\documentclass{ws-mpla-notrim}
\usepackage{amssymb,amsmath}
\def\bfnabla{\mbox{\boldmath $\nabla$}}

\def\bfsigma{\mbox{\boldmath $\sigma$}}
\def\lQ{\Lambda_{\rm QCD}}
\newcommand{\nn}{\nonumber}
\newcommand{\be}{\begin{equation}}
\newcommand{\ee}{\end{equation}}
\newcommand{\bea}{\begin{eqnarray}}
\newcommand{\eea}{\end{eqnarray}}

\def\als{\alpha_{\rm s}}
\def\siml{{\ \lower-1.2pt\vbox{\hbox{\rlap{$<$}\lower6pt\vbox{\hbox{$\sim$}}}}\ }} 
\def\tensor{\overleftrightarrow}

\begin{document}

\markboth{Joan Soto}
{Model Independent Results for Heavy Quarkonium}

%
%

\title{MODEL INDEPENDENT RESULTS FOR HEAVY QUARKONIUM}

\author{\footnotesize JOAN SOTO\footnote{Member of CER  Astrophysics, Particle Physics and Cosmology, associated with Institut de Ci\`encies de l'Espai-CSIC.
}}

\address{Departament d'Estructura i Constituents de la Mat\`eria, Universitat de Barcelona \\Diagonal 647, E-08028 Barcelona, Catalonia, Spain
\\
soto@ecm.ub.es}

\maketitle


\begin{abstract}
We review a number of results for the spectrum and inclusive decays of heavy quarkonium systems which can be derived from QCD under 
well controlled approximations. They essentially follow from the hierarchy of scales in these systems, which can be efficiently exploited using non-relativistic effective field theories. 
In particular, we discuss under which conditions non-relativistic potential models emerge as effective theories of QCD.

\keywords{Heavy Quarkonium; Effective Field Theories; Non-Relativistic QCD.}
\end{abstract}

\ccode{PACS Nos.: 14.40.Gx, 18.38.Bx, 18.38.Cy, 18.38.Lg}

\section{Introduction}	

Heavy quarkonium systems have played a prominent role in our current understanding of the Standard Model. Indeed, both 
charm and bottom quantum numbers were discovered through heavy quarkonium systems, $J/\psi$\cite{c}, the lightest vector charmonium, 
and $\Upsilon (1S)$\cite{b}, the lightest vector bottomonium, respectively. They have also been important for our understanding of QCD, 
the sector of the Standard Model which concerns strong interactions. Indeed, since the heavy quark masses $m$ are larger than 
$\lQ$, the typical hadronic scale, two important properties of QCD play a r\^ole in these systems, asymptotic freedom and confinement. 
On the one hand, asymptotic freedom explains the narrow width of the lower laying states\cite{af}. 
On the other hand, they are the closest objects in nature to two ideal static color sources, 
whose energy behavior at large distances serves as an order parameter for confinement (in the absence of light quarks)\cite{wilson}.

It was soon realized that, due to asymptotic freedom, for sufficiently heavy quark masses, heavy quarkonium systems 
should be similar to positronium and amenable for a weak coupling analysis\cite{af}. Unfortunately, actual charm and bottom masses 
turned out not to be sufficiently heavy as to allow to explain the observed spectrum in the weak coupling regime\cite{vl}. However, they 
appeared to be heavy enough as to allow for a good phenomenological description of the spectrum by means of simple non-relativistic 
potential models (see Ref. \refcite{review} for a review). What to take as the  potential 
 was, and still is, the main input of such models. The question then arose whether 
such potential could in principle be obtained from QCD if reliable non-perturbative techniques were at hand. Formulas were produced for it 
in terms of expectation values of Wilson loops (to be evaluated non-perturbatively) in a $1/m$ expansion 
up to order $1/m^2$, including spin dependent and velocity dependent terms\cite{wilson,potentials}. However, when one loop results for the
potential became available from direct QCD perturbative calculations\cite{contra}, it was realized that some of them were not correctly reproduced by the perturbative evaluation of the 
Wilson loop formulas.

Inclusive heavy quarkonium decays to light particles were calculated using a factorization hypothesis. Namely, the short distance 
annihilation process was computed in QCD at quark level, and the long distance (non-perturbative) effects were taken into account
by the wave function (or derivatives of it) at the origin, which was calculated using potential models or dropped from suitable ratios. However, again, when one loop 
results became available it was noticed that IR divergences appeared in some of the short distance calculations\cite{IR}.

The understanding of the IR divergences was possible due to the introduction of Non-Relativistic QCD (NRQCD)\cite{BBL}. It was shown that color octet operators, 
which are absent in potential models, were necessary to cancel the above IR divergences, and hence the factorization hypothesis 
used so far were wrong\cite{BBL0}. The long distance part needs not only wave functions at the origin but also matrix elements of color octet 
operators, which were not computable in terms of potential models. 

Thus the lesson seemed to be that potential models cannot 
incorporate all relevant features of QCD for heavy quarkonium systems.
However, a indication occurred that it may not be necessarily the case. If one recalculates the formulas for the QCD potential in 
terms of Wilson loops from NRQCD instead of directly from QCD, the discrepancies with the direct QCD calculation mentioned 
above disappeared, if the matching coefficients of NRQCD were calculated at one loop\cite{oakes}. 

One of the aims of this brief review is to illustrate that suitable potential models can indeed be regarded as effective theories of QCD, and hence 
totally equivalent to it, in a very particular kinematical regime, and, as such, NRQCD color octet operators have a precise 
representation in them. This produces a number of model independent results for the inclusive decay widths to light particles 
and for the NRQCD matrix elements. The second aim is to illustrate that in the weak coupling regime, which corresponds to a
 different kinematical situation, potential models are not an effective theory of QCD. This regime is well understood and a number 
of higher order calculations are available.

Before entering the issues above, let us mention that heavy quarkonium physics
is experiencing a revival. Recently, new states have been discovered and new processes have been measured, some compatible with theoretical expectations\cite{comp}, others not\cite{uncomp}, which is triggering theoretical research. We refer the reader to Ref. \refcite{qwg} for up-dates of the current status of the field, to Ref. \refcite{bali} for an extensive theoretical review and to Ref. \refcite{exp} for a recent experimental account.

\section{Heavy quarkonium as a non-relativistic system}
  
A system is called non-relativistic if, in the center of mass frame, the typical three-momentum of a particle $p$ is much smaller than its 
mass $m$. This implies that the non-relativistic energy $E:=\sqrt{m^2 +p^2}-m \sim p^2/m$
is much smaller than $p$. Hence a hierarchy of scales exist $m >> p >> E$, which may be exploited in order to simplify calculations. In addition other scales may also be important depending on the particular non-relativistic system. For heavy quarkonium, $\lQ$ is also important. In fact, it already enters in the definition of heavy quark, namely a quark whose mass fulfills $m >> \lQ$. Such a definition together with asymptotic freedom, which implies $\als (m) << 1$, suggests that heavy quarkonium, namely a heavy quark and a heavy antiquark (not necessarily of the same flavor), is indeed a non-relativistic system.

Rather than exploiting the inequalities  $m >> p >> E$ and $m >> \lQ$ in every individual calculation, it is more convenient to built 
effective field theories (EFTs), which implement them at the Lagrangian level. This is the approach we will follow.

\section{Non-Relativistic QCD}

The Non-Relativistic QCD (NRQCD) Lagrangian has the following aspect\cite{BBL}

\bea
&&{\cal L}_{\rm NRQCD}= \psi^\dagger \Biggl\{ i D_0 + \,\frac {{\bf D}^2}{2 m}
+ c_F\, g {{\bf \bfsigma \cdot B} \over 2 m}
+ c_D \, g { \left[{\bf D} \cdot, {\bf E} \right] \over 8 m^2}
+ i c_S \, g { {\bf \bfsigma \cdot \left[D \times, E \right] }\over 8 m^2}+\cdots \Biggr\} \psi \nn \\ &&\nn\\
&& + \chi^\dagger \Biggl\{ i D_0 - \, {{\bf D}^2\over 2 m_2} 
- c_F\, g {{\bf \bfsigma \cdot B} \over 2 m_2}
+ c_D \, g { \left[{\bf D \cdot, E} \right] \over 8 m_2^2}
+ i c _S \, g { {\bf \bfsigma \cdot \left[D \times, E\right] }\over 8 m_2^2}+\cdots  \Biggr\} \chi \nn
\\ &&\nn\\
&&
+ {f_{1}(^1S_0) \over m^2}O_{1}(^1S_0) 
+ {f_{1}(^3S_1) \over m^2}O_{1}(^3S_1) 
+ {f_{8}(^1S_0) \over m^2}O_{8}(^1S_0) 
+ {f_{8}(^3S_1) \over m^2}O_{8}(^3S_1) +\nn\\&&\nn\\
&& +{f_1({}^1P_1)   \over m^4}  O_1({}^1P_1)
+ {f_1({}^3P_{0}) \over m^4}  O_1({}^3P_{0})
+ {f_1({}^3P_{1}) \over m^4}  O_1({}^3P_{1})  
+ {f_1({}^3P_{2}) \over m^4}  O_1({}^3P_{2})
 + \cdots  \nn \\ && \nn \\
\eea
where
\bea
&&\nn\\
& O_{1}(^1S_0)= \psi^{\dag} \chi \chi^{\dag} \psi  \quad\quad , \quad\quad O_{1}(^3S_1) =\psi^{\dag} {\bfsigma} \chi \chi^{\dag} {\bfsigma} \psi & \nn \\ && \nn\\
& O_{8}(^1S_0) = \psi^{\dag} {\rm T}^a \chi \chi^{\dag} {\rm T}^a \psi  \quad\quad , \quad\quad 
O_{8}(^3S_1)= \psi^{\dag} {\rm T}^a {\bfsigma} \chi \chi^{\dag} {\rm T}^a {\bfsigma} \psi & \nn 
\\&&\nn\\
& O_1({}^1P_1) = \psi^\dagger (-\mbox{$\frac{i}{2}$} \tensor{\bf D}) \chi
        \cdot \chi^\dagger (-\mbox{$\frac{i}{2}$} \tensor{\bf D}) \psi &
\eea 

\bea
&
O_1({}^3P_{0}) =  {1 \over 3} \;
\psi^\dagger (-\mbox{$\frac{i}{2}$} \tensor{\bf D} \cdot \mbox{\boldmath $\sigma$}) \chi
        \, \chi^\dagger (-\mbox{$\frac{i}{2}$} \tensor{\bf D} \cdot \mbox{\boldmath
$\sigma$}) \psi &
\nn \\&&\nn\\
& O_1({}^3P_{1}) =  {1 \over 2} \;
\psi^\dagger (-\mbox{$\frac{i}{2}$} \tensor{\bf D} \times \mbox{\boldmath
$\sigma$}) \chi
        \cdot \chi^\dagger (-\mbox{$\frac{i}{2}$} \tensor{\bf D} \times
\mbox{\boldmath $\sigma$}) \psi &
\nn \\&&\nn\\
&
O_1({}^3P_{2}) = \psi^\dagger (-\mbox{$\frac{i}{2}$} \tensor{{\bf D}}{}^{(i}
\bfsigma^{j)}) \chi
        \, \chi^\dagger (-\mbox{$\frac{i}{2}$} \tensor{{\bf D}}{}^{(i} \bfsigma^{j)}) \psi &\nn .
\eea
$\psi$ is a Pauli spinor which annihilates a heavy quark and $\chi$ a Pauli spinor which creates a heavy antiquark. $c_F$, $c_D$, $c_S$, $f_1$, $f_8$, etc. are matching coefficients which encode (non-analytic) contributions from (relativistic) energy scales of order $m$, and may have a factorization scale ($\mu$) dependence.
The NRQCD Lagrangian is obtained from QCD by integrating out energy fluctuations about the heavy quark mass and three-momenta higher than, or of the order of, $m$ for the heavy quarks, and four momenta higher than, or of the order of $m$, in the gluon fields. This 
can be done in perturbation theory in $\als (m)$\cite{BBL} since $\als (m) << 1$ (see\cite{Manohar:1997qy,Pineda:1998kj} for an efficient 
way of doing such a calculation). Hence NRQCD is equivalent to QCD at any desired order in $\als (m) $ and $1/m$.  Note that the NRQCD Lagrangian is 
organized in inverse powers of $m$, which means that only the hierarchy
$m >>\lQ$, $p$, $E$ has been exploited. Hence, any dimensionful field in it does not have a definite size but may take the value of any of the remaining scales ($\lQ$, $p$, $E$). In spite of this, a concrete velocity ($v$) counting was put forward in
 the original papers under the assumption that $\lQ\sim E =: mv^2$ (then $p\sim mv$) which was useful to systematically organize calculations. As we will make clear in the following sections this is only one of the various counting possibilities 
that NRQCD admits, and may not be suitable for all heavy quarkonium states.   
In any counting, however, the scale dependence of the matching coefficients cancels
against the scale dependence induced by UV divergences in NRQCD calculations, and, hence, each $\mu$-dependence is eventually traded for one of the remaining dynamical scales ($\lQ$, $p$, $E$).   

Note that the NRQCD Lagrangian is manifestly invariant under rotations but not under Lorentz transformations. The Lorentz symmetry is, however, non-linearly realized and provides constraints on some of the matching coefficients (for instance, $c_S=2c_F-1$). These constraints were first uncovered in Ref. \refcite{Luke:1992cs} using reparametrization invariance. In Ref. \refcite{Brambilla:2003nt} it was shown that they follow form the Poincar\'e algebra.

The NRQCD Lagrangian contains non-hermitian terms due to imaginary pieces of the matching coefficients of the four quark operators ($f_1$, $f_8$,..)(see Ref. \refcite{Vairo:2003gh} for a recent update). This is due to the fact 
that a heavy quark and a heavy antiquark of the same flavor may annihilate into hard gluons (of energy $\sim m$), which have been 
integrated out. These non-hermitian pieces must be there in order to guarantee the equivalence of NRQCD and QCD. They contain 
crucial information about inclusive decay widths to light degrees of freedom. For instance, for P-wave states one obtains at leading 
order in the original velocity counting

\bea
&&\Gamma(\chi_Q(nJS)  \rightarrow {\rm LH})= 
{2\over m^2}\Bigg( {\rm Im \,}  f_1(^{2S+1}P_J) 
{\langle \chi_Q(nJS) | O_1(^{2S+1}P_J ) | \chi_Q(nJS) \rangle \over m^2}
\nn
\\
&&\quad
+ {\rm Im \,} f_8(^{2S+1}S_S) \langle \chi_Q(nJS) | O_8(^1S_0 ) | \chi_Q(nJS) \rangle\Bigg).
\label{NRQCDPW}
\eea 
where $\chi_Q(nJS)$ stands for a heavy quarkonium P-wave state of principal quantum number $n$, total angular momentum $J$ and spin $S$.
This is to be compared with the potential model result, which is recovered by dropping the second term and identifying

\be
\langle \chi_Q(nJS) | O_1(^{2S+1}P_J ) | \chi_Q(nJS) \rangle = {3 C_A\over 2 \pi} | R^{(0)\,\prime}_{n1} ({\bf 0}) |^2 
\ee
where $R^{(0)}_{n1} ({\bf r})$ is the wave function. The second term in (\ref{NRQCDPW}) is however crucial in order to cancel the scale ($\mu$) dependence of the matching coefficient of the first term at one loop  
\cite{BBL0}. For instance, in the $^3P_0$ case it reads\cite{Petrelli:1997ge}
\footnote{The $\mu$-independent piece of this result slightly differs from the earlier calculations in Refs. \refcite{IR}.}
\bea
&&
{\rm Im\,}f_1(^3P_0) 
= 3 C_F  \left( {C_A\over 2} -C_F \right) \pi \als(2m)^2
\left\{ 1 + {\als \over \pi}\left[ 
\left(-{7\over 3} + {\pi^2\over 4}\right) C_F \, + \right.\right.
\nn\\ && \left.\left. \qquad\qquad 
+ \left({427\over 81} - {\pi^2\over 144} \right) C_A 
+{4\over 27} n_f\left( -{29\over 6}  - \log {\mu \over 2m}\right) 
\right]\right\}.  
\eea

Let us also mention that the NRQCD formalism has also been used for the description of semi-inclusive decays (see Ref. \refcite{FL} and references therein) 
and inclusive production (see Ref. \refcite{bodwin} and references therein). We will not discuss these two applications here.

\section{Potential NRQCD}

NRQCD does not take advantage of the inequality $p >> E$. In particular, it contains gluons of typical energy $p$, which cannot
be produced in processes at the energy scale $E$. Simplifications should occur if one further integrates out
degrees of freedom with energies 
larger than $E$, 
which leads to Potential NRQCD (pNRQCD)\cite{Pineda:1997bj}.

Unlike in NRQCD, the degrees of freedom, and hence the Lagrangian, of pNRQCD depends on the interplay of $\lQ$ with $p$ and $E$. 
We shall discuss two situations below: the weak coupling regime ($k >> E \gtrsim \lQ$ ) and the strong coupling regime ($k \gtrsim \lQ >> E$), where $k$
 is the typical momentum transfer, which, for low lying states, is of the order of $p$.

\subsection{Weak coupling regime}

If $k >> E \gtrsim \lQ$ we can first integrate out energies $\sim k$ . The EFT thus obtained is pNRQCD in the weak coupling regime. It has the following aspect\cite{Pineda:1997bj,long}

\bea
\label{pnrqcdwc}
{\mathcal L}_{\rm pNRQCD} &=& 
{\rm Tr} \,\Big\{ {\rm S}^\dagger \left( i\partial_0 - h_s  \right) {\rm S} + 
+ {\rm O}^\dagger \left( iD_0 - h_o   \right) {\rm O} \Big\} + 
\nonumber \\ && \\
& & + \, {\rm Tr} \left\{\!   
{\rm O}^\dagger {\bf r} \cdot g{\bf E}\,{\rm S} 
+ \hbox{H.c.} + 
{{\rm O}^\dagger {\bf r} \cdot g{\bf E} \, {\rm O} \over 2} +
{{\rm O}^\dagger {\rm O} {\bf r} \cdot g{\bf E} \over 2} \!\right\} + \cdots
\nn
\eea
where $S=S({\bf R}, {\bf r}, t)$ and $O=O({\bf R}, {\bf r}, t)$ are singlet and octet wave function fields respectively, ${\bf R}$ is the center of mass coordinate (whose dynamics is trivial at lower orders and has been neglected above), ${\bf r}$ is the relative coordinate, and $h_s$ and $h_o$ are quantum mechanical Hamiltonians
\bea
\label{potentials}
h_s &=&  -{{\bf \nabla}^2  \over m}-C_f {\als \over r}+ \cdots
- \, {C_A \over  2}{\delta^{(3)}({\bf r})\over m^2}\Bigg ( 4 \, f_1(^1 S_0)
-2\,{\bf S}^2\left( f_1(^1 S_0)- f_1(^3 S_1) \right) \Bigg) + \nn\\
&&
+ {C_A\over m^4} \, {\cal T}^{ij}_{SJ} \bfnabla^i\delta^{(3)}({\bf r})\bfnabla^j \, 
f_1({}^{2S+1}P_J) + \cdots
\nn \\ && \\
h_o & = &  -{{\bf \nabla}^2  \over m} 
+  \left ( {C_A \over 2} - C_f \right ){\als \over r} + \cdots
\nn\\
&&
- \, {T_F \over 2}{\delta^{(3)}({\bf r})\over m^2}\Bigg ( 4 \, f_8(^1 S_0) -2\,{\bf S}^2\left( f_8(^1 S_0)-
  f_8(^3 S_1) \right) \Bigg) \nn 
\eea
${\cal T}^{ij}_{SJ}$ projects on states of spin $S$ and total angular momentum $J$ (see Ref. \refcite{Brambilla:2002nu} for a precise definition). 
Only gluons and heavy quarks of energies smaller than $k \sim 1/r$ are present in (\ref{pnrqcdwc}). However the only constraint on the three-momentum of the heavy quark is still that it must be smaller than $m$. The potentials in (\ref{potentials}) play the role of matching coefficients. They can be calculated perturbatively in $\als (k)$ ($\als (k) << 1$ since $k >> \lQ$) and in the $1/m$ expansion.
Beyond tree level, this calculation produces UV divergences and also IR divergences if the smaller scales ($E$, $\lQ$) are expanded. Once properly renormalized, the former cancel part of the scale dependence of the NRQCD matching coefficients. The latter and the remaining scale dependences from NRQCD matching coefficients cancel with scale dependences induced by properly renormalized UV divergences in pNRQCD calculations (see Refs. \refcite{lamb,cmy} for illustrations in QED) . 
Note also that if we drop the octet field
we recover a particular potential model. This is enough if 
we neglect non-perturbative contributions (meaning contributions for which the scale $\lQ$ plays a role), and are only interested in corrections up to $O(\als^2)$\cite{py,penin}. Beyond that 
order or if we wish to take into account non-perturbative contributions, the remaining  gluon and octet fields are crucial.

Note also that, analogously to NRQCD, the pNRQCD Lagrangian is manifestly invariant under rotations, but not under Lorentz transformations. The constraints from the full Poincar\'e algebra have been worked out in Ref. \refcite{Brambilla:2003nt}.

Let us mention here that, in spite of the fact that the top quark decays due to the weak 
interactions before forming hadronic states, this regime is also relevant for the study of 
the top-antitop system near its production threshold\cite{tt}.

\subsubsection{Spectrum}

The corrections to the spectrum up to order $\als^2$ had already been obtained
 before the introduction of pNRQCD\cite{py} ( and, in fact, making no use of NRQCD). 
The one loop potentials of the singlet field had been 
directly calculated from QCD\cite{contra}. The two loop potential, which was also necessary at this 
order, was calculated using static heavy quark propagators\cite{Schroder:1998vy}. 
NRQCD and pNRQCD just make the calculation simpler. Beyond 
that order or if one is interested in non-perturbative contributions due to the scale $\lQ$ all degrees of freedom of pNRQCD play a role and correct results cannot be obtained by just calculating  potentials
to a higher order\footnote{However, the leading\cite{vl} and next-to-leading\cite{p}  non-perturbative contributions in the case $E >> \lQ$  were obtained before the introduction of pNRQCD.}. In order to proceed further one has to specify the size of
$\lQ$ with respect to $E$. If $\lQ \sim E$, the leading non-perturbative effects are parameterized by non-local condensates and compete in size with the $\als^2$ perturbative corrections. If $E >> \lQ$, one can carry out weak coupling calculations with the (ultrasoft) gluons in (\ref{pnrqcdwc}). The physical observables can then be organized in powers of $\als$ (at different scales) since $p\sim k\sim m \als $ and $E \sim m{\als}^2$.
The logarithmic contributions to the corrections at $O(\als^3)$ were calculated in Ref. \refcite{logs} and the finite parts for the ground state 
in Ref. \refcite{hamburg}. Not only that, the use of EFTs, in this case NRQCD and pNRQCD, allows to resum IR QCD logarithms. For heavy 
quarkonium systems this was first proposed in Ref. \refcite{oakes} within NRQCD, later addressed in a slightly different EFT framework called 
vNRQCD\cite{vnrqcd}, and  implemented in the NRQCD-pNRQCD framework in Refs. \refcite{rgs,rgd}, which produced the first correct NNLL resummations 
for the complete spectrum\cite{rgs} (see also\cite{hs}). NLL resummation for the hyperfine splitting have been obtained recently\cite{hamburgp}. Non-perturbative contributions are parameterized in this case by local condensates\cite{vl}.

\subsubsection{Inclusive Decays}

The information on the parton subprocesses of inclusive decay widths to light particles (light hadrons, photons or leptons) is encoded in the imaginary parts of the NRQCD matching
 coefficients. These are inherited in pNRQCD as imaginary parts of local potentials ($\delta ({\bf r})$ and derivatives of it), which eventually 
makes the decay width proportional to the wave function at the origin (or derivatives of it). Corrections up to $\als^2$ to the wave
function at the origin were known before the introduction of pNRQCD\cite{penin}. In the case $E >> \lQ$, the double logarithmic corrections at $\als^3$ were already
obtained in this framework\cite{Kniehl:1999mx} and, today, the single logarithmic corrections at that order 
are also known\cite{Kniehl:2002yv}. The resummation of logs at NLL has already been carried 
out\cite{rgd,Pineda:be}. Semi-inclusive radiative decays have recently been addressed within this framework\cite{xj}.

\subsection{Strong coupling regime}

If $k \gtrsim \lQ >> E$, the integration of all degrees of freedom with energies larger than $E$ cannot be carried out in an expansion in $\als$.
In this case the degrees of freedom of pNRQCD are the singlet field interacting with a potential and the pseudo-Goldstone bosons\cite{long}.
If the latter are ignored, the form of the pNRQCD Lagrangian reduces to that of a potential model.

\be
{\mathcal{L}}_{\rm pNRQCD}= 
{\rm Tr} \,\bigg\{ {\rm S}^\dagger \left( i\partial_0 - h  \right) {\rm S} \bigg \} 
\label{pnrqcdsc}
\ee
where $h$ is a quantum mechanical Hamiltonian. Again, $h$ is manifestly invariant under rotations, but not under Lorentz transformations. The constraints of the full Poincar\'e invariance have been discussed in Ref. \refcite{Brambilla:2001xk}.

In the particular case $k >> \lQ $, we may integrate out first energies of the order of $k$, which can be done in an expansion in $\als (k)$ and $1/m$ exactly in the same way as in the weak coupling regime. We are thus lead to the same Lagrangian (\ref{pnrqcdwc}), which may be renamed as pNRQCD' since it is not our final EFT yet. We still have to integrate out energies $\sim \lQ$. This cannot be done in perturbation theory in $\als$  anymore but one can use the fact that $k$, $p >> \lQ >> E$. If one assumes that the octet field develops a gap $\sim \lQ$, it can be integrated out and we are left with the singlet field only. Namely we recover the degrees of freedom of a potential model as explicitely shown in (\ref{pnrqcdsc}).

In the general case $k \sim \lQ$, the integration of energies of order $k$ cannot be done perturbatively in $\als$. If one assumes that the potentials are analytic in $1/m$,  
\be
h=-{\bfnabla^2 \over m}+V_0+{V_1 \over m}
+{ V_2 \over m^2}+\cdots + { V_4 \over m^4}+ \cdots \,.
\ee
one can obtain them by matching (\ref{pnrqcdsc}) to NRQCD in the $1/m$ expansion. Then one obtains the non-perturbative potentials in terms of expectation values of 
operator insertions in  Wilson loops\cite{Brambilla:2000gk}. In this way one is able to reproduce and correct earlier results\cite{potentials}.  In particular,
it was noticed in this approach that the $1/m$ potential had been missed before.

\be
V_1(r) = \lim_{T\to\infty}\left(
- {g^2\over 4 T}\int_{-T/2}^{T/2} \!\! dt \int_{-T/2}^{T/2} \!\!dt^\prime
\vert t -t^\prime \vert \bigg[ \langle\!\langle {\bf E}(t) \cdot {\bf E}(t^\prime)\rangle\!\rangle_\Box 
- \langle\!\langle {\bf E}(t)\rangle\!\rangle_\Box \cdot
\langle\!\langle {\bf E}(t^\prime)\rangle\!\rangle_\Box  \bigg] \right).
\label{v1E}
\ee
where $\Box$ stands for a rectangle of size $T\times r$. $\langle\!\langle \cdots \rangle\!\rangle_\Box$  means the expectation value of the depicted fields joined by Wilson lines along the rectangle, divided by the expectation value of the Wilson loop in  the same rectangle. 

Let us make a parenthesis here and exemplify how the mismatch between the earlier Wilson loop approach and the explicit QCD one loop calculations mentioned in the introduction is resolved in the present formalism. Consider, for instance, one of the terms in $V_2$ contributing to the hyperfine splitting, 

\bea
V_2 & = & {{\bf S}^2}V_{S^2}^{(1,1)}(r)+ \cdots \nn \\ & & \\
V_{S^2}^{(1,1)}(r) & = & {2 c_F^2 \over 3}i \lim_{T\rightarrow \infty}\int_0^{T} dt \,  
 \langle\!\langle g{\bf B}_1(t) \cdot g{\bf B}_2 (0) \rangle\!\rangle_\Box 
- 4(d_{sv} 
+ d_{vv} 
C_f) \,
\delta^{(3)}({\bf r}) \nn
\eea  
where the $d_{sv}$ and $d_{vv}$ are suitable combinations of the $f_1(^{2S+1}S_S)$ and $f_8(^{2S+1}S_S)$ matching coefficients of the four fermion operators in (1)\cite{Pineda:1998kj}. In the earlier Wilson loop approach one would obtain the same expression with $c_F=1$ and $d_{sv}=d_{vv}=0$, namely the short distance contributions coming from loops and virtual annihilation processes from scales of the order of $m$ were missing. If one calculates the chromomagnetic correlator at one loop one finds a contribution proportional to $\als^2 \log (r \mu )$ which adds to a $\als^2 \log (m / \mu )$ contribution in $d_{sv}+ d_{vv}C_f$ producing the full QCD $\als^2 \log ( r m )$ contribution. 

Let us remark that the short distance behavior of these potentials
can be calculated in perturbation theory in $\als (r)$ and hence they must coincide with the ones in $h_s$ of the weak coupling regime (7). Therefore, they become increasingly singular at short distances as we go further in the $1/m$ expansion. Hence the Hamiltonian $h$ is not well defined in standard quantum mechanics. In order to make sense of it we must understand this Hamiltonian as an EFT. As such we should regulate it, establish a counting, and treat the subleading pieces as perturbations. The scale dependence induced by the regularization should cancel exactly with that in the NRQCD matching coefficients, much in the same way as it was observed in Ref. \refcite{cmy} for QED.    

The power counting in $h$ depends on the typical value of $p$ and $r$ in the concrete bound state we wish to analyze, and hence a simple power counting cannot be fixed a priori. Only a few statements can be made in general. The kinetic term and $V_0$ must always be assigned the same size ($mv^2$ since $p\sim mv$) and taken as leading order. Although $V_1$ is suppressed by $\als^2$ and hence order $mv^4$ in the weak coupling regime, it may in our case also be leading order, since in the strong coupling regime $\als \sim 1$ and dimensional counting allows for a size $V_1\sim (mv)^2$. The terms in $V_2$ are at most order $mv^3$ (although in the weak coupling regime are order $mv^4$ due to extra $\als$ suppressions) and hence they can be treated as perturbations.  

One should be aware that on general grounds the form of the potentials in (9) is not unique. Unitary transformations are allowed in quantum mechanics which change the aspect of the Hamiltonian but do not change the physics. In EFT one should better stick to transformations which preserve the counting. Even within those
one can reshuffle contributions from one term to another in the potentials\cite{Brambilla:2000gk}. In general different ways of performing the matching from NRQCD (or directly from QCD) lead to different forms of the potential related by unitary transformations. For instance, in the weak coupling regime, matching on-shell matrix elements rather than using the $1/m$ expansion, or matching in the Feynman gauge rather than in the Coulomb gauge produces different forms of the potential.

Recently, it has been shown that contributions to the potentials which are non-analytic in $1/m$ exists. A procedure to 
compute the ones due to the three-momentum scale $\sqrt{m\lQ}$ was put forward in Ref. \refcite{Brambilla:2003mu}. They give rise to subleading contributions
 with respect to the $1/m$ potentials. All these potentials (analytic and non-analytic) can be evaluated on the lattice\cite{Bali:1997am}.
Further non-analytic terms may appear due to the three-momentum scale $m\als$ when this scale is much larger than $\lQ$, which have not 
been taken into account so far.

\subsubsection{Spectrum}

Once the non-perturbative potentials are obtained from a lattice calculation (or by means of other non-perturbative methods\cite{Brambilla:1999ja,hagop}), one may think that the Schr\"odinger equation can be solved and the spectrum obtained\cite{Bali:1997am} in total analogy with potential models\cite{review}. A fully consistent calculation, however, requires the lattice calculation of the potentials to be translated to $\overline{MS}$ scheme or similar, in order to match the available NRQCD matching coefficients, or vice-versa. Furthermore, because of the same reason, one has to use (or to translate to) $\overline{MS}$ scheme the calculations in quantum mechanics perturbation theory. The advantage is that now one has 
a counting and a well-defined procedure which allows, 
at least in principle, to systematically improve the calculation by adding higher order potentials and by going to higher order in quantum mechanics perturbation theory. 
The contribution to the spectrum from potentials which are non-analytic in $1/m$ turns out to be very suppressed. 
The role of pseudo-Goldstone modes has not been addressed in this framework so far. 

\subsubsection{Inclusive decays}

As in the weak coupling regime, the imaginary parts of the NRQCD matrix elements are inherited in local terms of the 
pNRQCD Lagrangian. In particular, the local terms in $h_s$ (7) also exists in $h$. More important the color octet operators of NRQCD also have a representation in terms of local potentials. For instance, if we restrict ourselves to the $P$-wave contributions we have
\bea
&&
{\rm Im} \, V_4
=
C_A \, {\cal T}^{ij}_{SJ} \bfnabla^i\delta^{(3)}({\bf r})\bfnabla^j \, 
{\rm Im}\,f_1({}^{2S+1}P_J)
\\
\nn
&&\qquad\qquad
+{T_F \over 9}{\cal E}_3
{\bfnabla}\delta^{(3)}({\bf r}){\bfnabla}
\Bigg(
4\,{\rm Im} \,f_8(^1 S_0)
-2\,{\bf S}^2 \left({\rm Im} \,f_8(^1 S_0)-{\rm Im}\,f_8(^3 S_1)\right) 
\Bigg)+\cdots
\eea
Then the decay width of 
$P$-wave states to light hadrons at leading order now reads\cite{Brambilla:2001xy} 

\be
\Gamma(\chi_Q(nJS)  \rightarrow {\rm LH})= 
{C_A\over \pi}{| R^{(0)\,\prime}_{n1} ({\bf 0}) |^2 \over m^4}
\Bigg[ 3 \, {\rm Im\,}\,   f_1(^{2S+1}P_J) 
+ {2 T_F\over 3 C_A } {\rm Im\,} \,  f_8(^{2S+1}{\rm{S}}_S) \, {\cal E}_3 \Bigg]. 
\label{PWp}
\ee
where ${\cal E}_3$  is a non perturbative parameter defined as

\be
{\cal E}_3 = 
{1 \over N_c}\int_0^\infty dt \, t^3 \langle g{\bf E}(t)\cdot g{\bf E}(0)\rangle,
\ee

By comparing with (\ref{NRQCDPW}), one may rephrase (\ref{PWp}) as

\be
\langle \chi_Q(nJS)\vert O_8(^1S_0)\vert \chi_Q(nJS) \rangle
= {T_F\over 3}
{\vert R^{(0)\,\prime}_{n1}({\bf 0})\vert^2 \over \pi m^2} {\cal E}_3.
\ee
Namely, one is able to obtain NRQCD color-octet matrix elements in terms of (derivatives of) wave functions at the origin, which 
contain all flavor and principal quantum number dependence,  plus extra
universal (depending on $\lQ$ only) non-perturbative parameters. 
One can check in perturbation theory that the scale dependence of ${\cal E}_3$
cancels exactly the scale dependence of ${\rm Im\,}\,   f_1(^{2S+1}P_J)$ (5).
The unknown non-perturbative parameters together with the 
wave functions  at the origin may drop from suitable ratios. They may also be extracted from data. This allowed to put forward a 
prediction for bottomonium states
in terms of data extracted from charmonium\cite{Brambilla:2001xy}, which turned out to be in reasonable agreement with the experimental results
when they came out\cite{Cinabro:2002ji}. 

If all the states below threshold for bottomonium and charmonium were in the strong coupling regime, and if we restrict ourself to potentials which are 
analytic in $1/m$, one would obtain a reduction of unknown NRQCD matrix elements by roughly a factor of two\cite{Brambilla:2002nu} at order $1/m^4$ (i.e. LO for P-wave states and NLO for S-wave states). Non-analytic 
terms $\sim \sqrt{m\lQ}$ give rise to subleading contributions (provided that $\sqrt{m\lQ} >> m\als (\sqrt{m\lQ})$), which, however, may be of the same order as analytic corrections to the leading 
analytic contributions\cite{Brambilla:2003mu}. They would slightly increase the number of unknown matrix elements.

\section{Discussion}

So far we have discussed a theoretical framework with almost no reference to any heavy quarkonium state 
in nature. If we wish to apply it to a concrete state we have to first figure out whether this state belongs 
to the weak or strong coupling regime, if any. This is not easy to establish a priori, since what scale 
plays the role of $\lQ$ or the typical value of $k$ or even $E$, cannot be extracted directly from the 
 experimental observables. One may try the weak coupling regime first and check: (i) if the expansion in 
$\als$ shows good convergence and (ii) if the leading non-perturbative effects are small. These appear 
to be fulfilled by the $\Upsilon (1S)$ system, and to a lesser extend for the $B_c$ and $J/\psi$ 
\footnote{For this to be so one has to properly take into account the renormalon 
singularities\cite{renormalon,sumino}}. The constrain (ii) 
is very restrictive since non-perturbative effects in the weak coupling regime grow with a large power of the principal 
quantum number\footnote{In spite of this, there are indications that using the weak coupling regime 
for excited states in the $\Upsilon$ system may give reasonable results\cite{sumino}}. Hence, most likely any excited state does not belong 
to the weak coupling regime. This does not mean that they belong to the strong coupling regime, since 
there is also the possibility that pNRQCD does not apply to them . Indeed, the constrain $k\sim \lQ >> E$, 
does not allow states with $k\sim E \sim \lQ$. If we accept Heavy Quark Effective Theory counting rules\cite{neubert}, these are states close to, or beyond, the heavy-light 
pair production threshold. Hence neither the weak nor the strong coupling regime are in principle applicable to these states. In order to make things concrete, let us advance what
we believe to be a reasonable (although not entirely conservative) assignment. For the $\Upsilon$ system,
$\Upsilon (1S)$ and $\eta_b (1S)$ belong to the weak coupling regime, whereas the remaining states below 
heavy-light pair production threshold may well be considered in the strong coupling regime. For the $\psi$ system,
$J/\psi$ and $\eta_c (1S)$ seem to be in the border line between the weak and strong coupling regimen, the lowest
 lying $P$-wave states in the strong coupling regime, and the remaining states (including $\psi (2S)$ and $\eta_c (2S)$)
are either too close or beyond  the heavy-light pair production threshold, so most likely pNRQCD is not applicable.
For the $B_c$ system, the pseudoscalar and vector ground states may well be in the weak coupling regime\cite{noraa,hamburgp},
 the lowest lying $P$-wave states and the first radial excitation of the $S$-wave states in the strong coupling regime, whereas 
for the remaining states pNRQCD would not be applicable.

\section{Conclusions}

EFTs techniques allow to exploit efficiently the various hierarchies of scales appearing in heavy 
quarkonium. They simplify and systematize earlier approaches ( for instance, potential models) 
and are powerful enough to put forward new results, which stem from QCD under well controlled 
approximations. This is so both in the weak and the strong coupling regime. In the weak coupling regime a number 
of explicit calculations have been carried out at higher orders in $\als$. In the strong coupling regime 
  non-trivial results for the NRQCD matrix elements have been obtained. The phenomenological consequences 
of these results have not been fully exploited yet.

\section*{Acknowledgments}

Thanks are given to N. Brambilla, D. Eiras, A. Pineda and A. Vairo for enjoyable collaborations which led to many of the results presented here. Thanks are also given to A. Pineda for comments on the manuscript. We acknowledge financial support from a CICYT-INFN 2003 collaboration contract,
the MCyT and Feder (Spain) grant FPA2001-3598, the
CIRIT (Catalonia) grant 2001SGR-00065 and the network EURIDICE (EU)
HPRN-CT2002-00311.

\end{document}